\documentclass{aa}  

\usepackage{graphicx}
\usepackage{txfonts}
\usepackage{multirow}
\usepackage[T1]{fontenc}
\usepackage{ae,aecompl}
\pdfoutput=1

\usepackage{lscape}
\usepackage{color}
\usepackage{tabularx}

\usepackage{graphicx}   
\usepackage{amsmath}    
\usepackage{amssymb}    
\usepackage{makecell}

\begin{document} 

\title{Evolved eclipsing binary systems in the Galactic bulge: Precise physical and orbital parameters of OGLE-BLG-ECL-305487 and OGLE-BLG-ECL-116218.}
\titlerunning{Accurate stellar parameters of two evolved eclipsing binary systems.}
\author{K. Suchomska\inst{1}\thanks{E-mail: ksenia@camk.edu.pl(KS)}
\and D. Graczyk\inst{2}
\and C. Ga{\l}an\inst{1}
\and O. Zi{\'o}{\l}kowska\inst{1}
\and R. Smolec \inst{1}
\and G. Pietrzy{\'n}ski\inst{1}
\and W. Gieren\inst{3}
\and S. Villanova\inst{3}
\and M. G{\'o}rski\inst{1}
\and I. B. Thompson\inst{4}
 \and P. Wielg{\'o}rski\inst{1}
\and B. Zgirski\inst{1}
 \and P. Karczmarek\inst{3}
\and B. Pilecki \inst{1}
\and M. Taormina\inst{1}
  \and W. Narloch\inst{3}
  \and G. Hajdu\inst{1}
  \and M. Lewis\inst{1}
  \and M. Ka{\l}uszy{\'n}ski\inst{1}
  \and G. Rojas Garc{\'i}a\inst{1}}

\institute{Nicolaus Copernicus Astronomical Centre, Bartycka 18, 00-716 Warsaw, Poland
\and Nicolaus Copernicus Astronomical Centre, Rabia{\'n}ska 8, 87-100 Toru{\'n}, Poland
\and Departamento de Astronom{\'i}a, Universidad de Concepci{\'o}n, Casilla 160-C, Concepci{\'o}n, Chile
\and Carnegie Observatories, 813 Santa Barbara Street, Pasadena, CA 911101-1292, USA}

\date{Received date/ accepted date}




\abstract
{}
{Our goal is to determine, with high accuracy, the physical and orbital parameters of two double-lined eclipsing binary systems, where the components are two giant stars. We also aim to study the evolutionary status of the binaries, to derive the distances towards them by using a surface brightness--colour relation, and to compare these measurements with the measurements presented by the Gaia mission.}
{In order to measure the physical and orbital parameters of the systems, we analysed the light curves and radial-velocity curves with the Wilson--Devinney code. We used $V$-band and $I$-band photometry from the Optical Gravitational Lensing Experiment (OGLE) catalogue and near-infrared photometry obtained with the New Technology Telescope (NTT) equipped with the SOFI instrument. The spectroscopic data were collected with the High Accuracy Radial velocity Planet Searcher (HARPS) spectrograph mounted at the ESO 3.6m telescope and the Magellan Inamori Kyocera Echelle (MIKE) spectrograph mounted at the 6.5m Clay telescope.}
{We present the first analysis of this kind for two evolved eclipsing binary systems from the OGLE catalogue: OGLE-BLG-ECL-305487 and OGLE-BLG-ECL-116218. 
The masses of the components of OGLE-BLG-ECL-305487 are $M_1$ = 1.059 $\pm$ 0.019 and $M_2$ = 0.991 $\pm$ 0.018 $M_\odot$, and the radii are  $R_1$ = 19.27 $\pm$ 0.28 and $R_2$ = 29.99 $\pm$ 0.24 R$_\odot$. For OGLE-BLG-ECL-116218, the masses are $M_1$= 0.969 $\pm$ 0.012 and $M_2$= 0.983 $\pm$ 0.012 $M_\odot$, while the radii are $R_1$= 16.73 $\pm$ 0.28 and $R_2$= 22.06 $\pm$ 0.26 $R_\odot$. The evolutionary status of the systems is discussed based on the  \textsc{parsec} and \textsc{MIST} isochrones. The ages of the systems were established to be between 7.3-10.9 Gyr for OGLE-BLG-ECL-305487 and around 10 Gyr for OGLE-BLG-ECL-116218.
 We also measured the distances to the binaries. For OGLE-BLG-ECL-305487, $d$= 7.80 $\pm$ 0.18 (stat.) $\pm$ 0.19 (syst.) kpc and for OGLE-BLG-ECL-116218, $d$= 7.57 $\pm$ 0.28 (stat.) $\pm$ 0.19 (syst.) kpc. }
{}

\maketitle


\section{Introduction}
Understanding the physics and evolution of stars is one of the most important tasks in astronomy. Precise determination of the physical parameters of stars, such as mass, radius, luminosity, or metallicity, gives us an opportunity to better understand stellar structure and evolution, as well as the evolution of the galaxies that host these stars. 
It is worth mentioning that modern theories of stellar evolution are able to predict stellar parameters such as radius, mass, and metallicity with high accuracy at every evolutionary stage. However, those predictions were calibrated based mainly on main-sequence stars, giving us no grounds to think that they are also fully correct for evolved stars. Moreover, comparisons of increasingly precise observations with theoretical models of stellar evolution reveal inaccuracies in the models, implying that certain aspects, such as convection and overshooting, still need to be refined.

When it comes to testing stellar evolutionary models, the accuracy of the determination of stellar parameters is of primary concern, as models rely strongly on radius, mass, or effective temperature. As was noted by \citet{2010A&ARv..18...67T}, only parameters determined with precision better than 3\% can provide sufficiently strong constraints, and thus models with inadequate physics can be rejected. Well-detached eclipsing binary systems, where the components are evolved stars, give us an opportunity to derive the physical parameters of these stellar components with such precision, making these systems perfect candidates for constraining stellar models. 

Although detached double-lined eclipsing binaries (DEB SB2), where the components are two evolved stars, serve as a perfect testbed for stellar evolutionary models, they are not easy to detect. This is mainly due to their long orbital periods and relatively short eclipses. So far, only a few such systems in our Galaxy have been analysed \citep{2015MNRAS.451..651S, 2019A&A...621A..93S, 2015MNRAS.448.1945H, 2019MNRAS.484..451H}, and just over a dozen in both the Large and Small Magellanic Clouds \citep[hereafter LMC and SMC, e.g.][]{2013Natur.495...76P, 2019Natur.567..200P, 2018ApJ...860....1G}. 
 In this paper we present a detailed analysis of the photometric and spectroscopic data of two double-lined eclipsing binary systems from the OGLE catalogue, identified as OGLE-BLG-ECL-305487 and OGLE-BLG-ECL-116218 \citep[hereafter BLG-305487 and BLG-116218, respectively,][]{2016AcA....66..405S}. Both systems are composed of two evolved giant stars and are located in the Galactic bulge. Basic information about both systems can be found in Table~\ref{tab:basic}. We provide a set of physical and orbital parameters for both systems, determined with very high accuracy, as well as derived distances towards them. In Sect.~\ref{observations} we present details regarding the collected photometric and spectroscopic data. Section~\ref{analysis} provides information about the methods of the analysis and the modelling of both systems. Sections~\ref{evolution} and~\ref{distance} focus on the evolutionary status of the two binaries and the distances towards them, and in Sect.~\ref{summary} we present a summary of the obtained results.  
 
\begin{table} 
\caption{Basic data for OGLE-BLG-ECL-305487 and OGLE-BLG-ECL-116218.}
\centering
  \resizebox{\columnwidth}{!}{\begin{tabular}{ccc}
   \hline
  & BLG-ECL-305487 & BLG-ECL-116218\\ 
  \hline
  2MASS & J18061144-2614037 & J17484411-2228537\\
  GAIA DR3 & 4063353316189129728 & 4068973783492778624 \\
  $\alpha$&18:06:11.45 & 17:48:44.11\\
  $\delta$&-26:14:03.5 &-22:28:53.8 \\
  Period (d)& 176.66& 80.59\\
  $V$ (mag) $^1$&18.201 $\pm$0.01& 17.594 $\pm$ 0.005\\
  $I$ (mag) $^1$& 14.795 $\pm$ 0.005& 14.765 $\pm$ 0.009\\
  $J$ (mag) $^2$& 12.397 $\pm$ 0.031& 12.647 $\pm$ 0.053 \\
  $K$ (mag) $^2$& 11.017 $\pm$ 0.027& 11.368 $\pm$ 0.061 \\
  $J$ (mag) $^3$&12.322 $\pm$ 0.045&12.680 $\pm$ 0.029\\
  $K$ (mag) $^3$& 10.801 $\pm$ 0.034& 11.317 $\pm$ 0.026\\ \hline
  \multicolumn{2}{l} {$^1$ -OGLE-IV Johnson-Cousins filters} \\
  $^2$ - UKIRT \\
  $^3$ - 2MASS \\
   \end{tabular}}
  \centering
  \label{tab:basic}
\end{table}

\section{Observations}
\label{observations}

\subsection{Photometry}
The $V$-band and $I$-band photometric data used in our analysis were collected with the Warsaw 1.3 m telescope at Las Campanas Observatory in Chile during the third and fourth phases of the OGLE project \citep{2003AcA....53..291U, 2012AcA....62..219S, 2016AcA....66..405S}. 
A total of 89 and 1996 measurements in the $V$-band and $I$-band, respectively, were obtained for BLG-305487. The light curve coverage in the $I$-band was sufficient for use in further analysis, while the measurements in the $V$-band filter served only to determine the out-of-eclipse luminosity of this system in the $V$-band. 
As for BLG-116218, a total of 27 and 729 measurements were obtained in the $V$-band and $I$-band, respectively. Similar to the case of BLG-305487, the data coverage for the light curve in the $I$-band was sufficient for use in further analysis, while the data points in the $V$-band filter were only used to establish the out-of-eclipse luminosity of the system in this band.  

In our analysis we also used near-infrared $J$-band and $K$-band photometry that we obtained with the ESO NTT telescope at La Silla Observatory, equipped with the SOFI camera. For both systems we obtained seven epochs of near-infrared photometry, all taken outside of the eclipses. We followed the process of reduction and calibration of the data onto the UKIRT system presented in \citet{2009ApJ...697..862P}.  
The obtained infrared luminosities of the systems were later transformed onto the Two Micron All-Sky Survey (2MASS) system using the prescription given in \citet{2001AJ....121.2851C}.

\subsection{Spectroscopy}
The spectroscopic data used in our analysis were collected with the use of the Clay 6.5-m telescope at Las Campanas Observatory, equipped with the Magellan Inamori Kyocera Echelle (MIKE) spectrograph and with the ESO 3.6-m telescope at La Silla Observatory, equipped with the High Accuracy Radial velocity Planet Searcher (HARPS) spectrograph. To obtain high-resolution spectra with the MIKE spectrograph, we used the 5 x 0.7 arcsec slit, which gave us a spectral resolution of $\sim$42 000. As for the HARPS spectrograph, it was used in the EGGS mode, which resulted in a spectral resolution of $\sim$ 80 000. For BLG-305487 we collected five and six measurements with the MIKE and HARPS spectrographs, respectively. As for BLG-116218, we collected ten spectra in total, out of which seven were with the MIKE spectrograph and three with the HARPS spectrograph.

\section{Analysis and results}
\label{analysis}

We started our analysis by measuring the radial velocities of the components of the systems. We used the Radial Velocity and Spectrum Analyzer (RaveSpan) software  \citep{2015ApJ...806...29P, 2017ApJ...842..110P}, which uses the Broadening Function formalism \citep{1992AJ....104.1968R, 1999ASPC..185...82R}.  We selected the template spectra from the synthetic library  of \citet{2005A&A...443..735C}. RaveSpan was also used to estimate the spectroscopic luminosity ratios of the components of the binaries, which were later compared with the luminosity ratios that we obtained from the Wilson--Devinney code solution (hereafter WD code) version 2007  \citep{2007ApJ...661.1129V, 1971ApJ...166..605W, 1979ApJ...234.1054W, 1990ApJ...356..613W} in order to check for consistency in both results. RaveSpan is also equipped with a simple spectral disentangling mode with a spectral tomography method based on \citet{2006A&A...448..283G}, which we used to obtain spectra of the individual components. 

To derive the physical and orbital parameters of the analysed systems, we used the WD code, which is equipped with the automated differential correction (DC) optimising subroutine. This code allowed us to simultaneously solve the $I$-band light curve and radial-velocity curves of the investigated systems, which is essential when it comes to obtaining a consistent model of a binary. 

In addition to the WD code, we used the JKTEBOP code \citep{2004MNRAS.351.1277S, 2005MNRAS.363..529S}. JKTEBOP is equipped with a Monte Carlo (MC) error analysis algorithm, which allowed us to compare the errors of the determined parameters that we obtained directly from the WD code with those from the MC simulations.

\subsection{Radial velocities}
\label{sec:RV}
\subsubsection{BLG-305487}

We measured the radial velocities of the components of BLG-305487 with the RaveSpan code over the wavelength range of 5400--6800 \AA. The mask that we used excluded $H_\alpha$ and telluric lines. We noticed a systematic offset between the radial velocities of the components measured based on the MIKE and HARPS spectra, and therefore we decided to apply a 0.59 km s$^{-1}$ correction to the measurements from the MIKE spectra. The measured radial velocities of the components of BLG-305487 are presented in Table~\ref{tab:velocities1}.

We also checked whether the components of the analysed systems rotate synchronously. To perform this check, we measured their rotational velocities by fitting rotationally broadened profiles in RaveSpan and compared the measured velocities with the expected synchronous velocities. We performed the fitting procedure using four spectra obtained with the MIKE spectrograph, which gave us mean broadening values of $v_{M_1}$=9.45 $\pm$ 0.73 km s$^{-1}$ and $v_{M_2}$=13.57 $\pm$ 1.51 km s$^{-1}$ for the primary and secondary components, respectively.  

To estimate the correct $v$sin$i$, macro-turbulence velocities and instrumental profile contributions have to be taken into account \citep[as described in][]{2015MNRAS.451..651S}. 
We determined the macro-turbulence velocities to be $v_{mt1}$=1.45 km s$^{-1}$ and $v_{mt2}$=1.35 km s$^{-1}$ for the primary and secondary components, respectively. The instrumental profile was estimated to be $v_{ip}$=4.28 km s$^{-1}$, assuming the resolution of the MIKE spectrograph to be $R$ = 42 000. Therefore, the final values of the rotational velocities are $v_1$sin$i$ = 8.30 $\pm$1.73 km s$^{-1}$ and $v_2$sin$i$ = 12.80 $\pm$ 2.51km s$^{-1}$ for the primary and secondary components. 

We compared those values with the expected equatorial rotational velocities. We determined the expected rotational velocities to be $v_1$=6.90 km s$^{-1}$ and $v_2$ =8.59 km s$^{-1}$. We also compared them with the pseudo-synchronous radial velocities (as described in \citet{2019A&A...621A..93S}). The expected pseudo-synchronous velocities are equal to $v_{ps1}$=7.31 km s$^{-1}$ and $v_{ps2}$=9.10 km s$^{-1}$.  
We concluded that the primary component rotates synchronously within the margin of error. As for the secondary component, both synchronous and pseudo-synchronous velocities are consistent with the measured rotational velocity within the 1.5$\sigma$, and therefore it is difficult to clearly state whether or not this component rotates synchronously or pseudo-synchronously. 

\subsubsection{BLG-116218}

Radial velocities of the components of BLG-116218 were calculated over the wavelength range of 5400--6800 \AA~using a mask that excluded $H_\alpha$ and telluric lines. Similar to the case of BLG-305487, we noticed a systematic shift between the velocities calculated using the spectra obtained with the MIKE and HARPS spectrographs. Therefore, we applied a correction of -0.35 km s$^{-1}$ to the radial-velocity measurements from MIKE. The radial-velocity measurements for both components with the applied correction are presented in Table~\ref{tab:velocities2}.

Again, we checked if the components rotate synchronously. We fitted rotationally broadened profiles using six spectra obtained with the MIKE spectrograph. The mean values of the broadenings for both components were $v_{M_1}$ = 11.56$\pm$ 0.82 km s$^{-1}$ and $v_{M_2}$= 17.88 $\pm$ 0.9 km s$^{-1}$ for the primary and secondary components, respectively. We determined the macro-turbulence  velocities to be $v_{mt1}$=1.33 km s$^{-1}$ and $v_{mt2}$=1.30 km s$^{-1}$, and the instrumental profile was $v_{ip}$=4.28 km s$^{-1}$. We estimated the values of the rotational velocities, $v$sin$i$, of each component to be  $v_1\sin{i}$ = 10.65 $\pm$ 1.82 and $v_2\sin{i}$ = 17.31 $\pm$ 1.59 km s$^{-1}$. 
The expected equatorial rotational velocities for this system are $v_1$ = 10.52 km s$^{-1}$ and $v_2$ = 13.86 km s$^{-1}$, and therefore we concluded that the primary component rotates synchronously within the margin of error. As for the secondary component, the measurements are consistent within nearly 2$\sigma$.

\begin{table} 
\caption{Radial-velocity measurements of  BLG-305487. }
\centering
   \begin{tabular}{lrrr}
   \hline
  HJD & V$_1$ & V$_2$ &Instrument\\ 
  --2450000 & (km s$^{-1}$) & (km s$^{-1}$) &\\ \hline \hline
6102.70911 &  242.919    &  197.218     &  MIKE\\
6103.85689 & 242.287      &197.233      &  MIKE\\
6448.73854   & 242.073      & 196.961        & HARPS\\
6449.64147   & 242.546      & 197.407      &  HARPS\\
6463.69375   & 241.389      & 196.832      &  HARPS\\
6490.76392   & 225.973      & 212.901      &  MIKE\\
6529.52949   & 198.383      & 242.781     &  HARPS\\
6558.60223   & 200.348      & 240.489      &  MIKE\\
6560.55872   & 202.192      & 239.059      &  MIKE\\
7159.68935   & 241.761      & 196.984      & HARPS\\
7161.75015   & 241.990      & 196.522           &HARPS\\ \hline
  \end{tabular}
  \centering
  \label{tab:velocities1}
\end{table}

\begin{table} 
\caption{Radial-velocity measurements of  BLG-116218. }
\centering
  \begin{tabular}{lrrr}

   \hline
   HJD & V$_1$ & V$_2$ &Instrument\\ 
  --2450000 & (km s$^{-1}$) & (km s$^{-1}$) &\\ \hline \hline
6103.84693  &  81.014    & 126.158    & MIKE\\
 6449.81548   &128.320    &  78.296    &HARPS\\
 6463.67133   &129.018      &77.437     &HARPS\\
 6487.78655    &80.902     &125.313    &MIKE \\
 6490.73148    &76.634     &129.317   &MIKE\\
 6579.53466    &73.279     &133.253    &HARPS\\
 6932.53224   &126.547      &78.475   &MIKE\\
 7193.56026   &123.448      &83.663   & MIKE\\
 7211.57767    &84.096     &123.013     &MIKE\\
 7296.53175    &76.890     &128.738   &MIKE\\ \hline
  \end{tabular}
  \centering
  \label{tab:velocities2}
\end{table}

\subsection{Spectral disentangling and atmospheric analysis}
\label{spec}
To disentangle the spectra, we used the RaveSpan code, which is equipped with a simple spectral disentangling mode with a spectral tomography method based on \citet{2006A&A...448..283G}. As the method requires high signal-to-noise ratio (S/N) data, for both systems we used only the spectra obtained with the MIKE spectrograph. The disentangled spectra were then used to perform atmospheric analysis of the individual components. 

In order to perform this analysis, we used the {\sl{binary}} version of the `Grid Search in Stellar Parameters' ({\sl GSSP}) software package \citep{2015A&A...581A.129T}. The code uses the spectrum synthesis method by employing the {\sl S$_{YNTH}$V} local thermodynamic equilibrium-based radiative transfer code \citep{1996ASPC..108..198T}. The grid of {\sl{MARCS}} model atmospheres \citep{2008A&A...486..951G} interpolated into 0.1\,dex resolution in $\log{g}$ and [M/H], which is provided with the programme, is well suited for the analysis of the atmospheres of K-type giants in these binary systems. The adopted procedure was very similar to that in our previous work, where we calibrated surface brightness--colour relations based on eclipsing binary stars from the solar neighbourhood \citep{2022arXiv220807257G, 2021A&A...649A.109G}. During this analysis we used the highest S/N section of our spectra, in the range of 6000--6794\AA\,, but the spectral regions containing artefacts from imperfectly removed features from oxygen (O$_2$: $\lambda \sim$ 6270--6320\,\AA) and water (H$_2$O: $\lambda \sim$ 6470--6575\,\AA) molecules in Earth's atmosphere were skipped.

The free parameters we searched for were effective temperature ($T_{\rm{eff}}$), metallicity ($[$M/H$]$), micro-turbulent velocity ($\xi$), and projected rotational velocity ($V_{\rm{rot}} \sin{i}$).  Most of the input values of the parameters were taken according to the results of modelling with the WD code.  We started to search around the solar value of metallicity $[$M/H$]$.  Initial rotational velocities were set to the values corresponding to synchronous rotation.  The {\sl{GSSP--binary}} version uses the wavelength-dependent flux ratio ($f_{\rm i}$), which is calculated using the ratio of the components' radii ($r_{1}/r_{2}$) as obtained from the light curve fit using the WD code.  The surface gravities were not free parameters but were fixed to the values from the WD code solution. We also made additional fits with surface gravity as a free parameter to probe the precision of the surface gravity parameter as obtained from the WD method. The precision from the surface-gravity fit was two orders of magnitude worse than that provided by the dynamical solution.  The {\sl binary} version does not enable the calculation of macro-turbulent velocity ($\zeta$) due to degeneracy with rotational velocity ($V_{\rm{rot}}$) with which it is strongly correlated -- its values were adopted as estimated in Sect.\,\ref{sec:RV}.

\begin{figure}[ht]
    \centering
    \includegraphics[width=.5\textwidth]{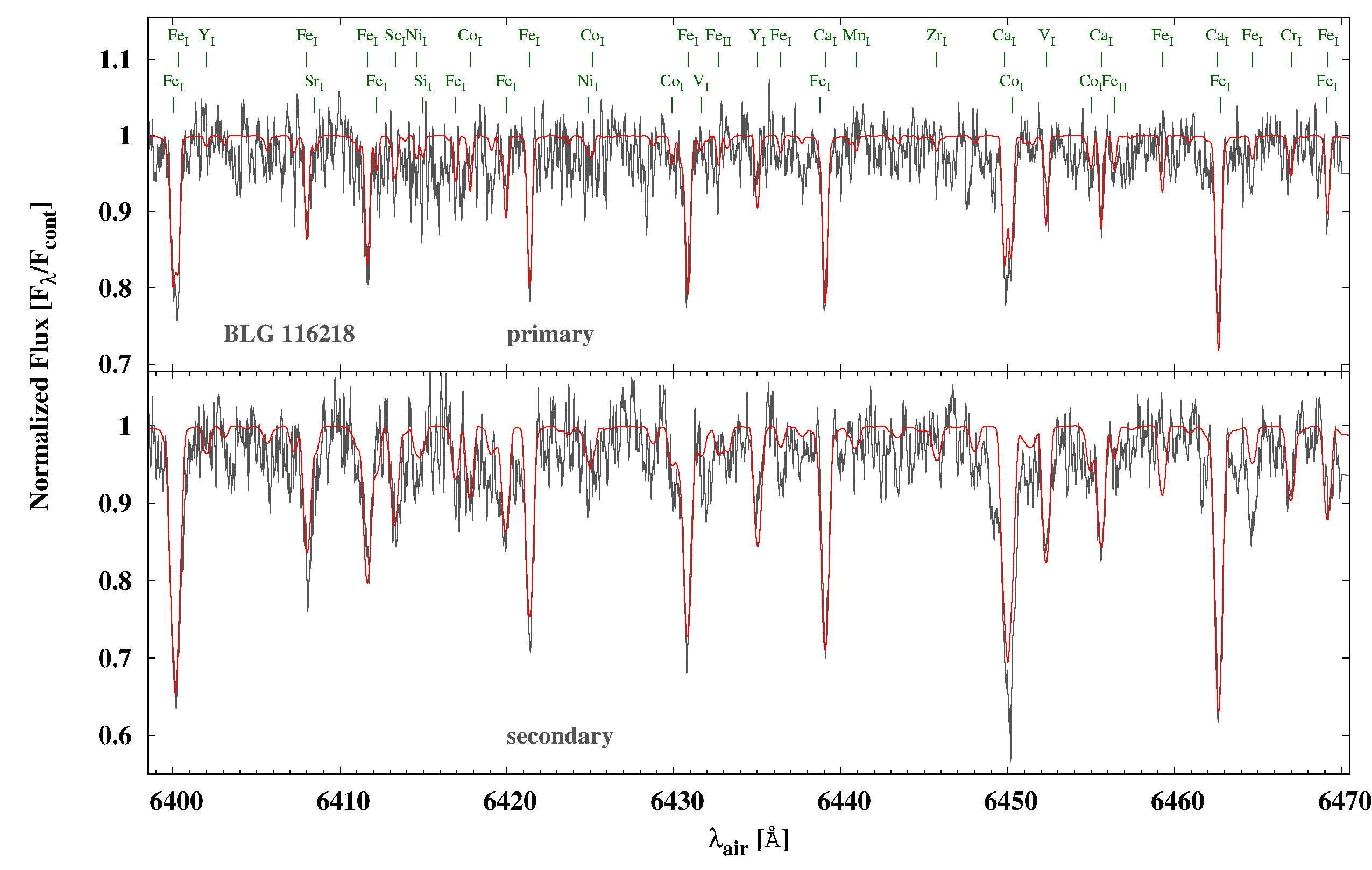}
    \caption{6399--6400\AA\, region of disentangled spectra (grey) of
    the primary (top) and the secondary (bottom) components of BLG-116218
    compared to the best-fit synthetic spectra (red).  Selected identified spectral features are shown at the top.}
    \label{fig:spectrum}
\end{figure}

Initially, relatively large steps in the grids of parameters were used to find the region close to the global minimum.  Then the parameter ranges were gradually narrowed and the sampling was made finer to find the solution corresponding to the best-matching model in three such iterations.  As an example of the analysis, in Fig.\,\ref{fig:spectrum} we show a section of the observed spectrum of BLG-116218  compared with the best-fit synthetic spectrum. The resulting final parameters are shown in Table\,\ref{tab.atmo1}.

\begin{table}
\caption{Atmospheric parameters of the components of BLG-305487 and BLG-116218.} 
\begin{tabular}{ccccc}
\hline
&\multicolumn{2}{c}{BLG-305487} & \multicolumn{2}{c}{BLG-116218} \\ 
& Prim. &Sec. &Prim. & Sec. \\  \hline \\
$T_{\rm{eff}}$ [K] &$4340^{+375}_{-440}$ & $3984^{+242}_{-254}$ & $4233^{+204}_{-196}$ & $4032^{+143}_{-135}$ \\ \\
 ${\rm [M/H]}$ & $0.1^{+0.53}_{-0.67}$ & $0.01^{+0.39}_{-0.34}$ &$-0.52^{+0.28}_{-0.33}$ &$-0.41^{+0.21}_{-0.24}$ \\ \\
log $g$ & $1.61^{+1.85}_{-1.80}$&$1.47^{+0.99}_{-0.89}$ & $1.95^{+1.09}_{-0.89}$&$1.73^{+0.48}_{-0.56}$ \\ \\ \hline 
\end{tabular}
\centering
\label{tab.atmo1}
\end{table}

\subsection{Interstellar extinction}

\subsubsection{BLG-305487}

In order to derive interstellar reddening towards BLG-305487, we used several calibrations of effective temperature - colour, $T_{\rm{eff}} - (V-K)$  \citep{1998A&A...339..858D, 1999A&AS..140..261A, 2000AJ....119.1448H, 2005ApJ...626..465R, 2006A&A...450..735M, 2009A&A...497..497G, 2010A&A...512A..54C}, where we used the effective temperatures established during the atmospheric analysis (see Table~\ref{tab.atmo1}). The determined reddening value was $E(B-V)$=1.4 $\pm$ 0.15 mag. The main contributions to the error come from the accuracy of the effective temperature $T_{\rm{eff}}$, the accuracy of the adopted effective temperature - colour calibrations themselves, and the accuracy of the V- and I-band photometry. 

We also estimated the reddening towards BLG-305487 using the extinction maps of \citet{1998ApJ...500..525S} with the recalibration of \citet{2011ApJ...737..103S}. A description of those calculations was presented in \citet{2015MNRAS.451..651S}. With this method, we estimated the reddening to be $E(B-V)$ = 1.59 $\pm$ 0.17 mag, assuming the distance to the system to be $d$ = 7.8 kpc. For the final reddening value, we decided to adopt the average of these two calculations $E(B-V)$ = 1.495 $\pm$ 0.179 mag, where the error is the combination of the statistical error and the systematic error.

\subsubsection{BLG-116218}

We used the same methods to estimate the reddening towards BLG-116218. From effective temperature $T_{\rm{eff}}$ -colour calibrations, we obtained $E(B-V)$ = 1.07 $\pm$ 0.1 mag. As for the Schlegel maps, the value was $E(B-V)$ = 0.808 $\pm$ 0.049 mag, assuming the distance to the system $d$ = 7.57 kpc. For the final value, we adopted the average of those two estimations $E(B-V)$ = 0.939 $\pm$ 0.111 mag, where the estimated error is the combination of the statistical error and the systematic error. 

\subsection{Light and radial-velocity curve modelling}

\subsubsection{BLG-305487}

To perform the modelling of both systems, we used the WD code, which allows us to simultaneously fit the multi-band light curves and radial-velocity curves in order to obtain a consistent model of a binary. We adopted the moment of primary minimum, $T_{0}$ = 2457028.84 days, and orbital period, $P$ = 176.66 days, from the OGLE catalogue. The orbital period together with the phase shift parameter (PSHIFT) were adjusted later in the analysis. As described in \citet{2019A&A...621A..93S}, PSHIFT allows a subroutine in the WD code to adjust for a zero-point error in the ephemeris used to compute the phases. The out-of-eclipse magnitudes for the $V$- and $I$-band filters were measured from all of the observational data out of minima and their mean values were $V$=18.201 and $I$=14.795 mag. In this work, by $primary$ $component$ we mean the star that is being eclipsed during the deeper, primary minimum. 

To perform the modelling, we used the WD code, fitting simultaneously the $I$-band light curve and the radial-velocity curve. The light curve coverage of the $V$-band photometry was not sufficient to include in the WD analysis, and therefore we only used those data to derive the out-of-eclipse magnitude. The input parameters for the DC subroutine were chosen as described in \citet{2012ApJ...750..144G}. To find the best-fit model of a binary, it is also crucial to decide which parameters are being adjusted. We chose the adjustable parameters, as well as  gravity brightening parameters and albedo, and adopted limb-darkening law in the same way as was described in \citet{2019A&A...621A..93S,2015MNRAS.451..651S}. The limb-darkening coefficients were computed according to the logarithmic law by \cite{1970AJ.....75..175K}. The coefficients were calculated internally by the WD code (setting LD = -2) during each iteration of the DC subroutine using tabulated data computed by \cite{1993AJ....106.2096V}. Models calculated using the linear and square root limb-darkening law, as well as models where coefficients of the linear limb-darkening law were treated as adjustable parameters, did not show any considerable improvement in the fit to the light curves. The parameters of gravity brightening and albedo were set to 0.32 and 0.5, respectively, which according to \cite{1967ZA.....65...89L} are appropriate values for the types of stars we are analysing.

The initial value for the temperature of the primary component was set to $T_{1}$= 5500 K. We used this only as a starting point in the DC subroutine of the WD code iterations. During this procedure all of the free parameters were adjusted and the best-fit model was found. We additionally checked for the presence of third light ($I_{3}$) to establish its impact on the final solution. The third light corrections were constantly negative, suggesting an unphysical solution. Therefore, we set $I_{3}$=0 in our final solution. 
Our next step was to compare the luminosity ratios of the components obtained from the WD code with the spectroscopic luminosity ratios obtained with the use of the RaveSpan code, in order to check if our model is in agreement with the corresponding spectroscopic information. We determined the spectroscopic luminosity ratio to be $\sim$ 1.78 for $\lambda$=6470$\AA$ , which is slightly lower than the luminosity ratio obtained from the WD code for the same wavelength. When both potentials of the components ($\Omega_{1}$ and $\Omega_{2}$) are set as adjustable parameters, the preferable light curve solution leads to a disagreement between the luminosity ratio and the spectroscopic luminosity ratio obtained with the RaveSpan code. Therefore, we decided to readjust the values of $\Omega_{1}$ and $\Omega_{2}$, and fix the value of the potential of the primary component. Readjusting means finding the closest value of $\Omega$ (of the primary or secondary component) that allows us to obtain a luminosity ratio that is in agreement (within the margin of error) with the spectroscopic luminosity ratio. In this way we obtained consistency between our spectroscopic and photometric solutions.  

After disentangling the spectra and performing the atmospheric analysis (see Sect. \ref{spec}), we obtained a better estimate of the temperatures of the components (Table~\ref{tab.atmo1}). We implemented the new values into the WD code and repeated the fitting procedure. The errors of the determinations of $T_{1}$ were significantly higher than those for the secondary component. This was caused by the much lower S/N in the spectra of the primary component. We decided to adjust the value of $T_{1}$ during the fitting procedure while the value of $T_{2}$ was fixed to the value adopted from the atmospheric analysis. We once again ran the DC subroutine of the WD code in order to get a new best-fit model. 

\begin{figure*}
\centering
\includegraphics[width=\textwidth]{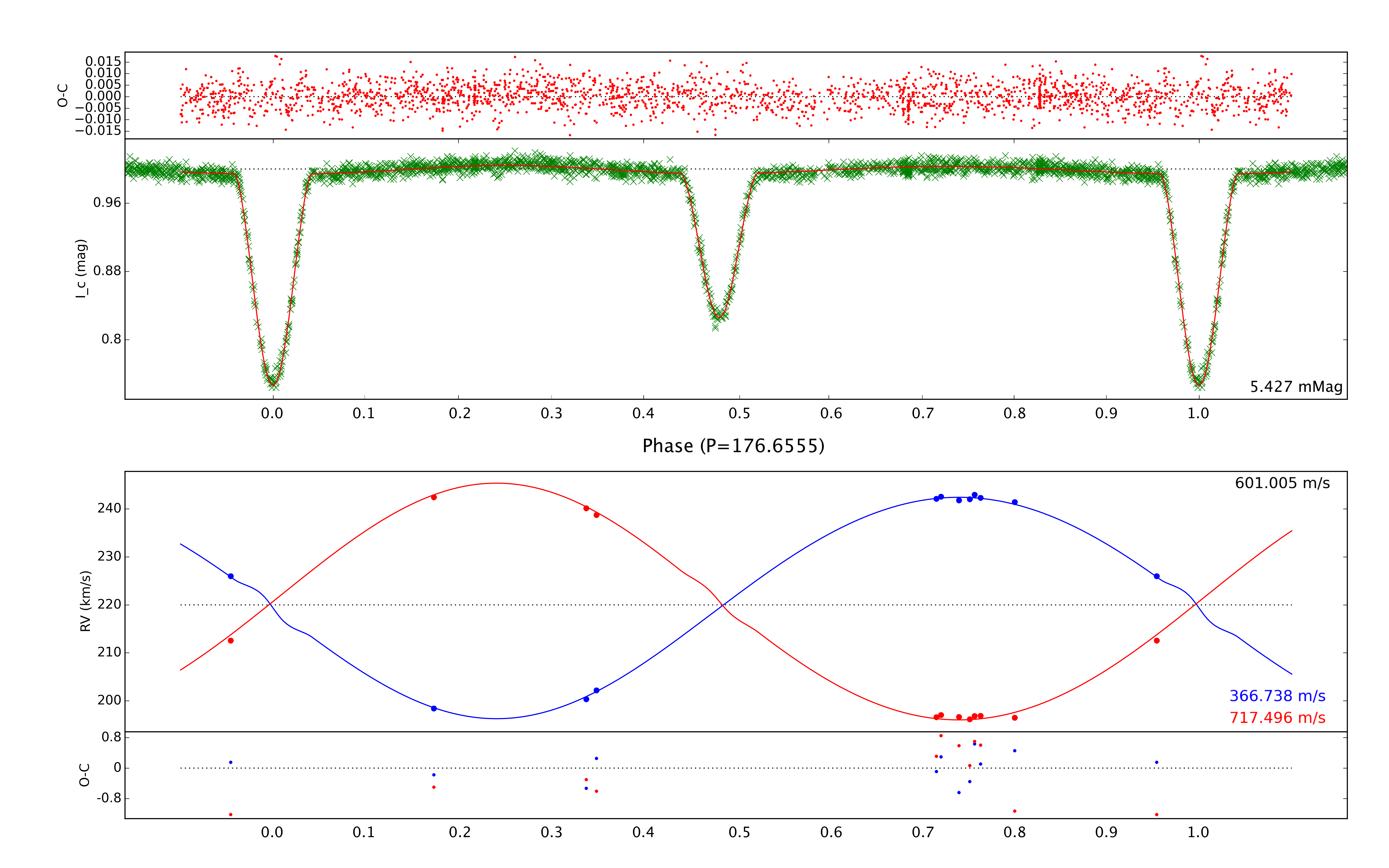}
    \caption{$I$-band light curve and radial-velocity solutions from the WD code for OGLE-BLG-ECL-305487. The blue circles represent the measurements of the primary component, while the red circles represent the measurements of the secondary component. The residuals of the fits are listed at the right ends of the panels. }
                \centering
    \label{fig:blg305487_wd}
\end{figure*}

The $I$-band light curve together with the radial-velocity solution are presented in Fig.~\ref{fig:blg305487_wd}. The parameters derived from the WD code are presented in Table~\ref{table:results1}.  

\subsubsection{BLG-116218}
The moment of primary minimum and orbital period were initially adopted from the OGLE catalogue, and their values were $T_{0}$ = 2457041.51 and $P$ = 80.59 days. Both of these parameters were later adjusted during the fitting procedure. We also measured out-of-eclipse magnitudes in both the $V$- and $I$-band filters, and their values were 17.594 and 14.765 mag, respectively. As before, the primary component is the one being eclipsed in the deeper primary minimum.

To perform the modelling, we once again used the WD code. We simultaneously fit the $I$-band light curve and the radial-velocity curve. As before, the $V$-band photometry data were not sufficient to cover the light curve, and therefore we used them only to determine the out-of-eclipse magnitude. The choice of gravity brightening parameters, albedo, and adopted limb-darkening law was the same as fo the case of BLG-305487. As for the adjustable parameters, the eccentricity ($e$=0), argument of periastron ($\omega_{0}$), and PSHIFT were not adjusted this time. The orbit of the system is circular as we cannot see any systematic deviations in the light curves or radial velocities, which would have been the case when adjusting a model of a circular orbit to an orbit that is actually eccentric. 

We set the initial value of temperature of the primary component, $T_{1}$, to 5500 K. This was our starting point for the iterations with the DC subroutine of the WD code and the preliminary best-fit model was found. At the end of this procedure, we also checked for the presence of third light $I_{3}$. Again, the results continuously suggested an unphysical solution, and therefore we fixed the value to $I_{3}$ = 0. 

We also compared the spectroscopic luminosity ratio obtained with the RaveSpan code with the photometric luminosity ratios ($L_{2}/L_{1}$) obtained from the WD solution. We determined the spectroscopic luminosity ratio for $\lambda$=6470$\AA$ to be $\sim$ 1.43. To obtain an agreement between these two determinations, we readjusted the values of the potentials of the components ($\Omega_{1}$ and $\Omega_{2}$) in the WD code and then decided that $\Omega_{2}$ was no longer going to be adjusted in further analysis. 

\begin{figure*}
\centering
        \includegraphics[width=\textwidth]{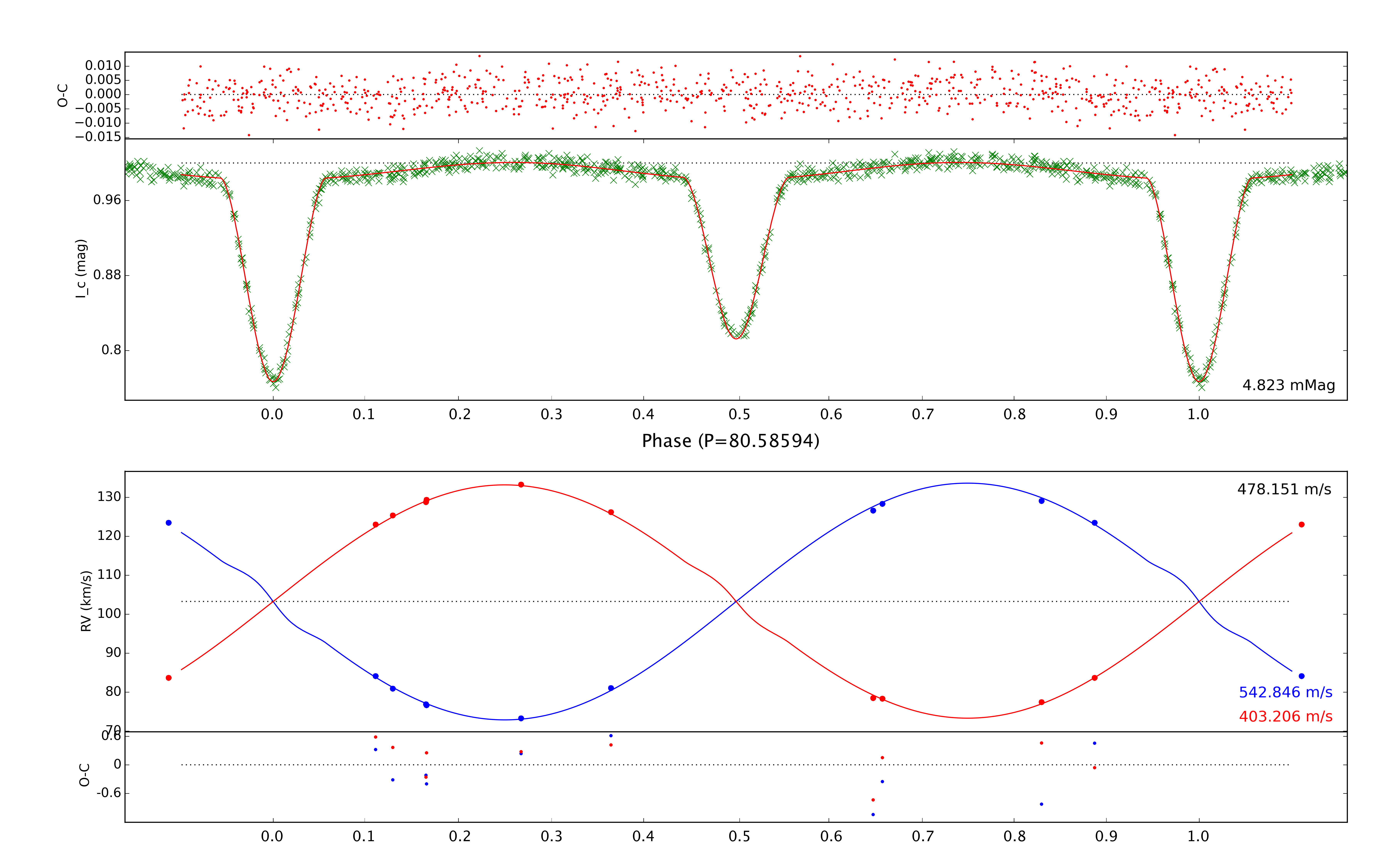}
    \caption{Same as for Fig.\ref{fig:blg305487_wd}, but for OGLE-BLG-ECL-116218.}
                \centering
    \label{fig:blg116218_wd}
\end{figure*}

The spectral disentangling and atmospheric analysis procedures provided us with better determinations of temperatures of both components (see Table~\ref{tab.atmo1}). We used these improved temperatures in the WD code and repeated the fitting procedure with the temperature of the secondary component being adjusted. After several iterations with the DC subroutine of the WD code, we found the final solution of our model. 
The $I$-band light curve together with the radial-velocity solution are presented in Fig.~\ref{fig:blg116218_wd}. The derived parameters from the WD code are presented in Table~\ref{table:results1}.

\begin{table*} 
\caption{Photometric and orbital parameters obtained with the WD code for the BLG-305487 and BLG-116218 systems.}
\centering
 \begin{tabular}{lll}
\hline
Parameter & BLG-305487 &BLG-116218 \\ \hline \hline
Orbital inclination  $i$ (deg) & 82.03 $\pm$ 0.02& 78.60 $\pm$ 0.07 \\
 Orbital eccentricity $e$ & 0.0283 $\pm$ 0.0007& 0 (fixed)\\
Primary temperature $T_1$ (K) &   4248$\pm$ 6&   4233 (fixed)\\
Secondary temperature $T_2$ (K)& 3984 (fixed) & 4083$\pm$ 5\\
Fractional radius $r_1$ & 0.1145 $\pm$ 0.0015 & 0.1705 $\pm$ 0.0028\\
Fractional radius $r_2$ &  0.1783$\pm$ 0.0009&  0.2247$\pm$ 0.0025\\
$k = r_2/r_1$ & 1.5568 $\pm$ 0.0219& 1.3180 $\pm$ 0.0261\\
Orbital period $P$ (day) & 176.6549 $\pm$ 0.0015& 80.5859 $\pm$ 0.0014\\
$\omega$ (deg) & 181.4 $\pm$ 7.2& 62.23(fixed)\\
$\Omega_1$ & 9.7033 (fixed) & 6.8984 $\pm$ 0.0447\\
$\Omega_2$ & 6.3269 $\pm$ 0.0486&  5.5398 (fixed) \\
$(L2/L1)_V$ &  1.4519 &  1.3003\\
$(L2/L1)_I$ & 1.7102$\pm$ 0.0666 & 1.4239$\pm$ 0.0689\\
$(L2/L1)_J$ & 1.9670& 1.5488\\
$(L2/L1)_K$ & 2.2176& 1.6596\\
$T_0$ (JD-2450000)& 7028.84 (fixed) & 6235.624$\pm$ 0.011\\
 Semi-major axis $a$ (R$_\odot$) & 168.370$\pm$ 0.998 & 98.168$\pm$ 0.369\\
 Systemic velocity $\gamma$ (km s$^{-1}$) & 219.99 $\pm$ 0.12& 103.23 $\pm$ 0.45\\
Velocity semi-amplitude  $K_1$(km s$^{-1}$) & 23.08 $\pm$ 0.15& 30.41 $\pm$ 0.16\\
Velocity semi-amplitude  $K_2$(km s$^{-1}$) & 24.68 $\pm$ 0.24& 29.99 $\pm$ 0.16\\
Mass ratio $q$ & 0.935 $\pm$  0.009& 1.014 $\pm$  0.008\\ 
RV $rms_1$  (km s$^{-1}$) &0.373&0.543\\
RV $rms_2$  (km s$^{-1}$) &0.764&0.403\\ \hline
 \end{tabular}
 \label{table:results1}
\end{table*}

\subsection{Absolute dimensions}
We present the final values of the physical parameters for the components of both BLG-305487 and BLG-116218  in Table~\ref{tab:results_final1}. In order to derive the masses of the components, we followed the equation:
\begin{equation}
M_1[M_\odot] = 1.34068 \times 10^{-2} \frac{1}{1+q}\frac{a^3[R_\odot]}{P^2[d]},
\end{equation}
\begin{equation}
M_2 [M_\odot] = M_1 \times q,
\end{equation}
where $q$ is the mass ratio, $a$ is the semi-major axis, and $P$ is the orbital period. 
 
We derived the observed magnitudes of each component and their colour relations (see Table~\ref{tab:results_final1}) using the equations:
\begin{equation}
\label{V1}
V_1=V - 2.5 \log{ \frac{1}{1+\frac{L_2}{L_1}}},
\end{equation}

\begin{equation}
\label{V2}
V_2= V - 2.5\log{ \frac{\frac{L_2}{L_1}}{1+\frac{L_2}{L_1}}},
\end{equation}
where $V_1$ and $V_2$ are the magnitudes of the primary and secondary components, and $\frac{L_2}{L_1}$ is the luminosity ratio in a given filter. Similarly, we determined the magnitudes for the $I$-, $J$-, and $K$-band filters. The $V$-band magnitudes were converted to bolometric magnitudes using the bolometric corrections from \citet{1999A&AS..140..261A}.

\begin{table*}
\caption{Physical properties of the BLG-305487 and BLG-116218 systems.}
\centering
  \begin{tabular}{lllll}
  \hline
  &\multicolumn{2}{c}{BLG-305487}&\multicolumn{2}{c}{BLG-116218}\\
  Property & The Primary &The Secondary & The Primary& The Secondary\\ \hline 
  Spectral type & K0 III & K2 III & K2 III & K3 III\\
  $V$ (mag) & 19.175 &  18.770& 18.498 &  18.213\\
  $I$ (mag) & 15.878 & 15.295& 15.726 & 15.343 \\
  $J$ (mag) & 13.713 & 12.979 & 13.795 & 13.320\\
  $K$ (mag) & 12.292 &  11.427& 12.442 & 11.892\\
  $V\!-\!I$ (mag) &  3.297 & 3.475&  2.772 & 2.871  \\
  $V\!-\!K$ (mag) & 6.883 & 7.343& 6.056 &  6.321 \\
  $J\!-\!K$ (mag) & 1.422 &  1.552& 1.353 & 1.427 \\
  Radius ($R_{\odot}$) & 19.27$\pm$ 0.28 & 29.99 $\pm$ 0.24& 16.73$\pm$ 0.28 & 22.06 $\pm$ 0.26 \\
  Mass ($M_{\odot}$) & 1.060 $\pm$  0.020& 0.991 $\pm$ 0.018 & 0.969 $\pm$  0.012& 0.983 $\pm$ 0.012\\
  log $g$  & 1.893 $\pm$ 0.013 & 1.480 $\pm$ 0.005 & 1.978 $\pm$ 0.015 & 1.744 $\pm$ 0.009 \\
  $T_{\rm eff}$ (K) &$4248_{-440}^{+375 *}$& $3984_{-254}^{+242 **}$& $4233_{-196}^{+204 *}$ & $4083_{-135}^{+143 **}$ \\
  $v$ sin $i$ (km s$^{-1}$) & 7.41 $\pm$ 1.73 & 12.42 $\pm$  2.3 & 10.65 $\pm$ 1.82 & 17.31$\pm$  1.59 \\
  Luminosity ($L_{\odot}$) & 108 $\pm$ 41&203  $\pm$ 51 & 81 $\pm$ 15&121  $\pm$ 17\\
  $M_{\rm bol}$ (mag) & -0.383 & -1.132 &0.018 & -0.444\\
  $M_{\rm v}$ (mag) &0.373 & -0.032 &0.786 & 0.501 \\
  Bolometric correction & -0.695 $\pm$ 0.045 & -1.053 $\pm$ 0.063 & -0.703$\pm$ 0.045 & -0.883 $\pm$ 0.054 \\
  $[$Fe/H$]^{**}$ & $0.1_{-0.67}^{+0.53}$& $0.01_{-0.34}^{+0.39}$& $-0.52_{-0.33}^{+0.28}$ & $-0.41_{-0.24}^{+0.21}$\\ \hline
  $E(B\!-\!V)$ & 1.495$\pm$ 0.179  && 0.939$\pm$ 0.111&\\
  Distance (pc) & 7803.7$\pm$179.9(stat.)& $\pm$  198.3  (syst.) & 7569.8$\pm$275.3(stat.) & $\pm$  189.1  (syst.)\\ \hline
 $^{* -\rm WD\: solution}$ \\
 $^{**- \rm atmospheric\: analysis}$\\
\end{tabular}
\centering
\label{tab:results_final1}
\end{table*}

\section{Evolutionary status}
 \label{evolution}
 
Modelling evolved binary systems is a challenging task as many factors need to be taken into account. These include, for example, overshooting at the border of the convective core, mass loss, rotation of the two components, and rotational mixing, as well as any dynamical interaction of the two components during evolution. Dedicated studies indicate that the inclusion and fine-tuning of some of these processes may be essential to correctly describe the system's evolution and model its present state (e.g. \cite{2017MNRAS.468.3533E}).
 Such a detailed treatment is beyond the scope of the present paper; to asses the evolutionary status of our systems, we used the much simpler approach of isochrone fitting, which still provides useful constraints on the evolutionary status of our systems and is a good starting point for further, dedicated modelling.

To determine the evolutionary status of the eclipsing binary systems, we used the publicly available {\sc{parsec}} (version 1.1) isochrones \citep{2012MNRAS.427..127B}. We downloaded isochrones with $9.3 \leq {\rm log}(age) \leq 10.2$, with a step of 0.001\,dex. Metallicity was varied from its base value (mean of the spectroscopic metallicities of the two components) by $\pm 0.2, \pm 0.4 \, \rm dex$.

To find the optimal age and metallicity of the systems, we applied a $\chi^2$ minimisation procedure as follows. First, for a given isochrone in our grid we calculated the $\chi_{\rm p,\,s}^2$ for the primary and secondary components separately, which included effective temperature, luminosity, radius, and mass, for each mass point on the isochrone in the $M\pm \sigma(M)$ range, where $\sigma(M)$ is a 1$\sigma$ error on the mass determination. Then, we calculated the final $\chi^2$ value for the isochrone as $\chi^2 = \chi^2_{\rm p,\,min} + \chi^2_{\rm s,\,min}$. Finally, this was repeated for all the downloaded isochrones to obtain the best solution for a given system.

We investigated the effect of red-giant branch (RGB) mass loss on our solutions. The adopted Reimers' formula \citep{1985iue..prop.2255R} is characterised by a free parameter $\eta_{\rm R}$. Asteroseismic calibrations based on old open clusters \citep{2012MNRAS.419.2077M} suggest that  $\eta_{\rm R}$ should be set in the range $0.1 \lesssim \eta_{\rm R} \lesssim 0.3$. Initially, three commonly used literature values were considered: $0.0$ (i.e., no mass loss), $0.2$, and $0.4$. With this procedure, we conducted our calculations for three grids of isochrones for each of the two binary systems.

The two systems have slightly different mass ratios. In the case of BLG-116218, the grids had to be slightly extended, which we discuss below. 

\subsection{BLG-305487}
\begin{figure*}[ht]
    \centering
    \includegraphics[width=.8\textwidth]{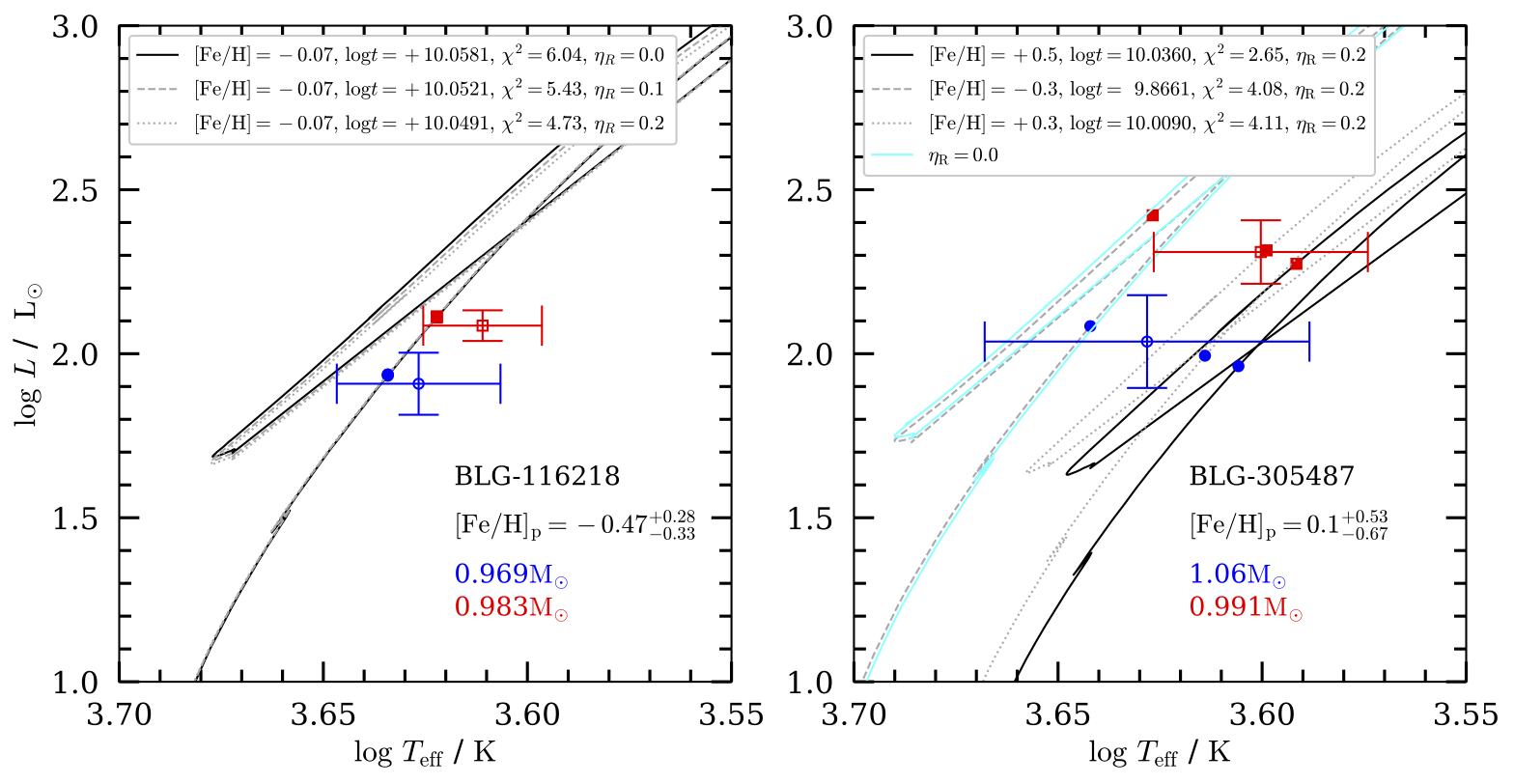}
    \caption{Three {\sc parsec} isochrones for BLG-116218 and BLG-305487 with the lowest $\chi^2$ in our grid. The parameter $\eta_{\rm R}$ quantifies the mass-loss rate on the RGB. The cyan line represents tracks with no mass loss. The primary star is marked with blue circles, while the secondary is marked with red squares. Open and filled symbols correspond to the observed and modelled values, respectively. Metallicity, age, and $\chi^2$ of the individual isochrones are given in the upper left corner.}
    \label{fig:hr_new}
\end{figure*}

\begin{figure*}[ht]
    \centering
    \includegraphics[width=\textwidth]{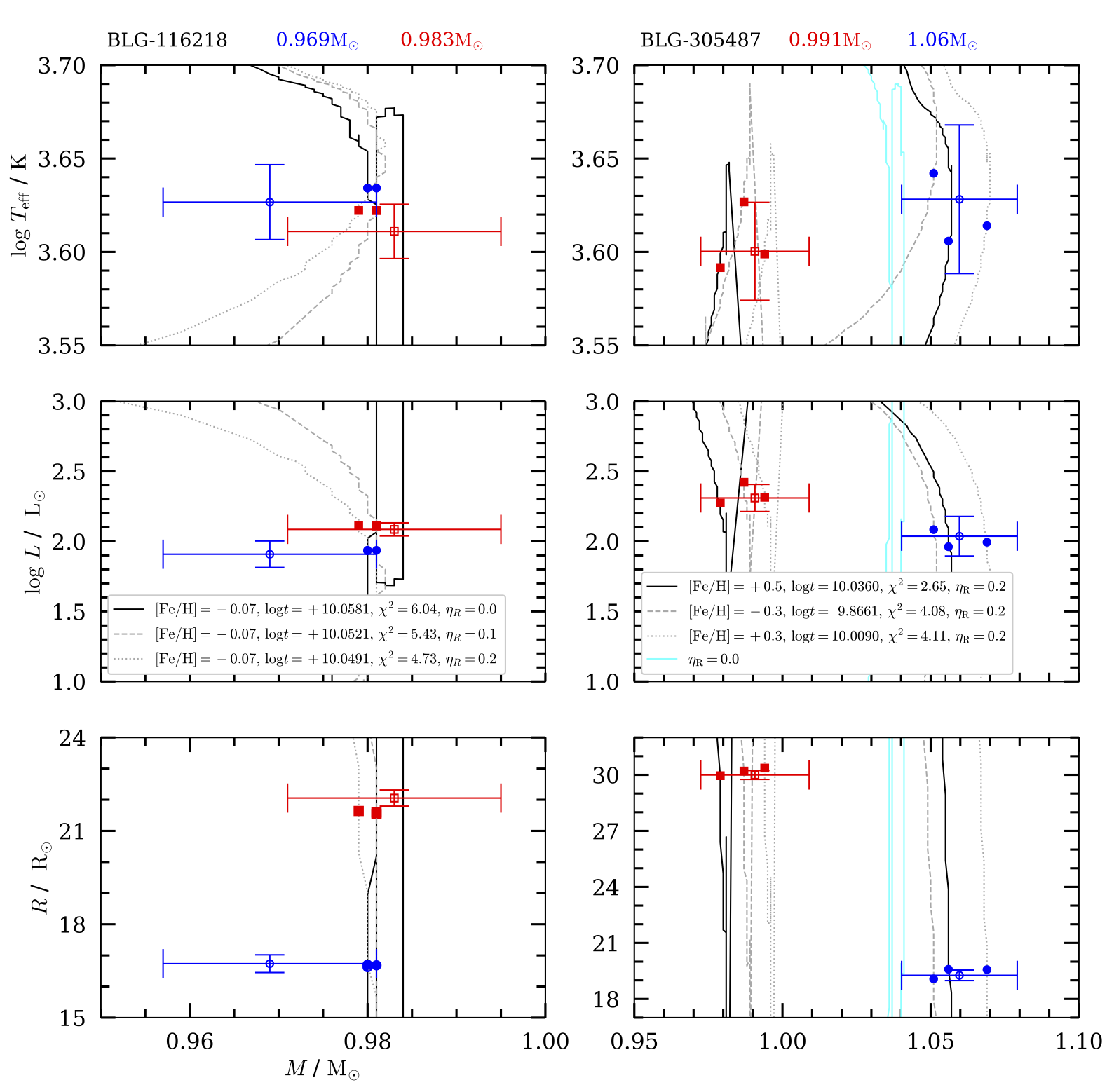}
    \caption{Same as for Fig.~\ref{fig:hr_new}, but as a function of mass.}
    \label{fig:iso_new}
\end{figure*}

Our three best solutions for BLG-305487, which differ in metallicity, are presented on a Hertzsprung-Russell (H-R) diagram (Fig.~\ref{fig:hr_new}) and in Fig.~\ref{fig:iso_new}, with effective temperature, luminosity, and radius each plotted as a function of mass. The age of the system is between $\sim 7.3$ and $\sim10.9 \rm \, Gyr$. The uncertainty of the spectroscopic metallicity is high and we find good solutions across a wide range of metallicities, from $-0.3$ to $0.5$ dex.

We were not able to find a good solution for a high mass-loss rate, nor for the case without mass loss. In the latter case, we show the exemplary isochrone with cyan lines. In Fig.~\ref{fig:iso_new}, this no-mass-loss isochrone runs vertically (through a very narrow mass range) through three evolutionary stages: RGB ascent, descent after helium ignition, and asymptotic giant branch (AGB) ascent, from right to left. This imposes very similar masses of binary stars during these evolutionary stages. With moderate mass loss applied, these vertical sections become broader in mass and a proper solution can be found. The primary star is ascending on the RGB while the secondary is on the AGB, as indicated by the lack of helium in the centre.

\subsection{BLG-116218}
For this system the mean spectroscopic metallicity and its uncertainty are $-0.47 \pm 0.26 \, \rm dex$. 
The isochrones with the mean spectroscopic metallicity lie parallel to the observations, on their hot side. This could be explained by a systematic shift in the effective temperatures of about $400 \rm K$ from the spectroscopic values. Another explanation could lie in the underestimation of the metallicity. We found a good solution for BLG-116218 for $\rm[Fe/H] = -0.07 \, dex$, which is within $\pm 0.4 \, \rm dex$ from the mean spectroscopic value.

As the components of BLG-116218 have a lower mass ratio than BLG-305487, the system requires slower mass loss to reproduce their location on the mass diagrams. We investigated values between 0.0 and 0.2, and the best results were obtained for $\eta_{\rm R} = 0.2$, which still agrees with asteroseismic calibrations. We present our best solution for three different mass-loss rates in Figs.~\ref{fig:hr_new} and ~\ref{fig:iso_new}. The resulting age of this system is $\sim 10 \rm Gyr$. Both components are ascending on the RGB.

\subsection{MIST isochrones}
The $\chi^2$ minimisation  procedure was repeated for the MIST \citep{2016ApJ...823..102C, 2016ApJS..222....8D} isochrones that are based on the {\sc{mesa}} evolutionary code (\citet{MESA-I, MESA-II, MESA-III, MESA-IV}). The mass-loss rate in MIST is calibrated to fit various observational constraints and the user cannot modify it when downloading the isochrones. On the RGB, the mass-loss rate in MIST is computed assuming Reimers' formula with $\eta_{\rm R}=0.1$. The resulting evolutionary stage and age of the stars agree with those derived from {\sc{parsec}}. The fit for BLG-116218 is slightly better using MIST rather than {\sc{parsec}} isochrones; while in the case of BLG-305487, the MIST fit is worse -- the isochrones on mass diagrams are too narrow, indicating a need for faster mass-loss rate, which agrees with the conclusion derived from {\sc{parsec}}.

\begin{table*}
 \centering
  \caption{Error budget of the distance moduli of BLG-305487 and BLG-116218.}
  \resizebox{\textwidth}{!}{ \begin{tabular}{@{}ccccccccccc@{}}
  \hline
  System & Type of error & $(m - M)$ & $\sigma$A   & $\sigma$diBenedetto & $\sigma E(B-V)$  & $\sigma V$ & $\sigma K$  & $(L_2/L_1)_K$ & Combined error  \\
&    &(mag)&(mag)&(mag)&(mag)&(mag)&(mag)&(mag) &(mag) \\ \hline \hline
  BLG-305487&   \textbf{Statistical}&14.462 & 0.012& --& $0.035^1$& 0.002& 0.032& -&\textbf{0.049}\\
   &\textbf{Systematic} & 14.462&--& 0.044 &--&\multicolumn{2}{c}{0.03}& 0.01&\textbf{0.054} \\   \hline
   BLG-116218&\textbf{Statistical}&14.395 & 0.009& --& $0.024^1$& 0.002 & 0.074& -&\textbf{0.078}\\
    &\textbf{Systematic} & 14.395&--& 0.044 &--&\multicolumn{2}{c}{0.03}& 0.01&\textbf{0.054} \\
        \hline
  \multicolumn{3}{c}1 - combination of statistical and systematic error
  \end{tabular}}
  \centering
  \label{tab:error1}
\end{table*}

\begin{table*}
\renewcommand{\arraystretch}{3}
\centering
\caption{ Comparison of distance determinations based on the surface brightness-colour relation, photometric parallaxes from those measurements, parallaxes from Gaia~eDR3, and distance determinations from \citet{2021AJ....161..147B} for five binary systems, where the components are giant stars analysed in this paper and by \citet{2015MNRAS.451..651S,2019A&A...621A..93S}.}
\begin{tabular}{|c|l|l|l|l|l|}
\hline
System & \makecell{$d$ (pc) \\ Suchomska et al.} & \makecell{$\varpi_{phot}$ (mas)$^1$ \\ Suchomska et al.} & \makecell{$\varpi$ (mas) \\ Gaia DR3} & \makecell{$d$ (pc) \\ Bailer-Jones \\geometric} & \makecell{$d$ (pc) \\ Bailer-Jones \\photogeometric}\\ \hline \hline
ASAS-180057$^2$ & \makecell{1600.5 $\pm$ 43.9(stat.)\\ $\pm$ 39.3(syst.)}& 0.6248 $\pm$ 0.0222& 0.5641 $\pm$ 0.0193& 1662.8$_{-62.5}^{+45.0}$ &1650.9$_{-52.9}^{+46.4}$ \\ \hline
BLG-123903 & \makecell{2953 $\pm$ 59.(stat.)\\ $\pm$70.2(syst.)}& 0.3386 $\pm$ 0.0102 & 0.2819 $\pm$ 0.0370 & 2952.4$_{-269.0}^{+331.9}$ & 2714.4$_{-200.0}^{+399.8}$ \\ \hline
BLG-296596 & \makecell{5682.8 $\pm$ 74.2(stat.) \\$\pm$142.3(syst.)} &  0.1759 $\pm$ 0.0048& 0.1437 $\pm$ 0.0307 & 5527.2$_{-732.9}^{+1030.5}$ & 5298.5$_{-637.5}^{+709.7}$\\ \hline
BLG-305487 & \makecell{7803.7 $\pm$ 179.9(stat.) \\$\pm$198.3 (syst.)}& 0.1281 $\pm$ 0.0042 & 0.1413 $\pm$ 0.0580 & 6460.1 $_{-1645.8}^{+2304.1}$ & 6861.2$_{-1414.1}^{+1833.8}$ \\ \hline
BLG-116218 & \makecell{7569.8 $\pm$ 275.3 (stat.) \\$\pm$189.1 (syst.)} & 0.1321 $\pm$ 0.0056& 0.1325 $\pm$ 0.0505 &  5876.0$_{-1102.4}^{+1528.3}$ & 5622.7$_{-896.5}^{+963.4}$ \\ \hline
\multicolumn{3}{l}{$^1$- \rm statistical\: and\: systematic\: error\: combined}\\
\multicolumn{3}{l}{$^2$ -\rm distance \: has \:been \: recalculated\: in\: this\: work}\\
\end{tabular}
\label{tab:odleglosci_all}
\end{table*} 

\section{Distances to BLG-305487 and BLG-116218}
\label{distance}
The distances towards the systems were determined using the surface brightness--colour relation for late-type stars established by \citet{2005MNRAS.357..174D}, analogous to the work presented in \citet{2019A&A...621A..93S}. Measured distances are  $d$=7803.7 $\pm$ 179.9(stat.)$\pm$198.3 (syst.) pc and $d$=7569.8 $\pm$ 275.3 (stat.)$\pm$189.1 (syst.) pc for BLG-305487 and BLG-116218, respectively. In Table~\ref{tab:error1} we present the main contributions to both the statistical error and the systematic error for the distance measurements for the analysed systems.   

We compared our distance determinations with those provided by the Gaia Data Release 3 (DR3). According to \cite{2022arXiv220800211G}, the parallaxes for our analysed systems are $\varpi$=0.1413 $\pm$ 0.0580 for BLG-305487 and $\varpi$=0.1325 $\pm$ 0.0505 mas for BLG-116218. The parallaxes determined based on our measured distances are $\varpi_{phot}$=0.1281 $\pm$ 0.0042 and $\varpi_{phot}$=0.1321$\pm$ 0.0056 mas for BLG-305487 and BLG-116218, respectively. The BLG-116218 parallax measurements seem to be in better agreement than those of BLG-305487. Nevertheless, for both systems, the determinations are comparable within the margin of error.

We also compared our results with the distance determinations presented by \citet{2021AJ....161..147B}, which are based on parallaxes in the Gaia early Data Release 3 (eDR3) \citep{2021A&A...649A...1G}. The authors made a catalogue of geometric distances for 1.47 billion stars and photogeometric distances for 92$\%$ of them. While the geometric distances use only parallaxes from the Gaia eDR3, the photogeometric ones also use the G magnitude and BP-RP colour from the eDR3. According to the authors, their estimates involve direction-dependent priors constructed from a sophisticated model of the 3D distribution, colours, and magnitudes of stars in the Galaxy as seen by Gaia. \citet{2021AJ....161..147B} estimate the distance towards BLG-305487 to be $d$=6460.1 $_{-1645.8}^{+2304.1}$ pc for the geometric measurement and $d$= 6861.2$_{-1414.1}^{+1833.8}$ pc for the photogeometric one. As for BLG-116218, the determined values are $d$=5876.0$_{-1102.4}^{+1528.3}$ pc and $d$=5622.7$_{-896.5}^{+963.4}$ pc for the geometric and photogeometric measurements, respectively. Both of our distance measurements are significantly higher than those presented by \citet{2021AJ....161..147B}. While the distance determinations for BLG-305487 are in agreement in the upper margin of error, for BLG-116218 we find better agreement when comparing Gaia parallaxes with the photometric parallaxes obtained from our measurements.  

In Table~\ref{tab:odleglosci_all} we present the distance determinations to five detached eclipsing binary systems, where the components are giant stars. The analysis of these binaries was presented in detail in this paper, as well as in \citet{2015MNRAS.451..651S, 2019A&A...621A..93S}. 

The distance measurement to ASAS-180057 differs from the one presented in \citet{2015MNRAS.451..651S} as we did not account for the fact that  the $2$MASS $J$- and $K$-band measurements were taken during the minimum. We recalculated the values provided by $2$MASS and established the out-of-eclipse magnitudes in these filters to be $J$=6.805 and $K$=5.369 mag. We also recalculated the reddening towards this system and obtained $E(B-V)$=0.672 $\pm$ 0.037 mag. As for the distance moduli, the recalculated value is $(m-M)$ = 11.021 $\pm$ 0.059 (stat.) $\pm$ 0.053 (syst.) mag. 

By comparing our distance measurements with those calculated by \citet{2021AJ....161..147B}, we note that with larger distances, the discrepancies between the measurements get bigger. The method of distance determinations presented in this section serves as an independent way of testing the parallaxes and distances provided by the Gaia mission.

\section{Summary and conclusions}
\label{summary}
We have presented the analysis of two detached double-lined eclipsing binary systems in the Galactic bulge, where the components are giant stars. We determined the physical and orbital parameters of these systems with an accuracy of between 1 and 2 \%. For our effective temperature estimations, the accuracy was significantly lower, especially for BLG-305487 due to the quality of the collected spectra and the low S/N. 

Using a surface brightness-colour relation, we measured the distance towards each of the systems and derived $d$ = 7803.7 $\pm$ 179.9(statistical error)$\pm$198.3 (systematic error) pc for BLG-305487 and $d$=7569.8 $\pm$ 275.3(statistical error)$\pm$189.1 pc (systematic error) for BLG-116218. The accuracy of our determinations is at the level of $\sim$ 3\%, where the main contribution to the error comes from the interstellar extinction and the uncertainties of the infrared photometry collected for these systems. We compared our results with those provided by the Gaia mission and presented the comparison in Table~\ref{tab:odleglosci_all}, along with the measurements for three other evolved binary systems analysed by our group \citep{2015MNRAS.451..651S, 2019A&A...621A..93S}.

Determining physical parameters of evolved stars with an accuracy below 2\% is very important for testing stellar evolution theory. So far only a few binary systems in our Galaxy, where the components are giant stars, have been studied with such precision \citep[e.g.][]{2015MNRAS.448.1945H}. Stellar evolutionary models are sensitive to many parameters, and therefore analysing and providing increasingly precise determinations of physical properties of evolved systems is much needed.

\begin{acknowledgements}
We would like to thank the staff of the Las Campanas and ESO La Silla Observatories for their support during the observations. 
We gratefully acknowledge financial support for this work from the Polish National Science Centre grant MAESTRO 2017/26/A/ST9/00446. The research leading to these results has received funding from the European Research Council (ERC) under the European Union’s Horizon 2020 research and innovation program (grant agreement No. 951549) and from the Polish National Science Center grants: MAESTRO 2017/26/A/ST9/00446 and BEETHOVEN 2018/31/G/ST9/03050. We also acknowledge the grant MNiSW DIR/ WK/2018/09. W.G. and G.P. gratefully acknowledge financial support for this work from the BASAL Centro de Astrofisica y Tecnologias Afines (CATA) AFB-170002. W.G. also gratefully acknowledges support from the ANID BASAL project ACE210002. OZ and RS acknowledge support from the Polish National Science Centre grant SONATA BIS 2018/30/E/ST9/00598.

This work is based on observations collected at the European Organisation for Astronomical Research in the Southern Hemisphere under ESO programmes: 095.D-0026(A), 092.D-0363(A). We are pleased to thank the ESO,  Chilean, and Carnegie Time Assignment Committees for their generous support of this program.
This work has made use of data from the European Space Agency (ESA) mission {\it Gaia} (\url{https://www.cosmos.esa.int/gaia}), processed by the {\it Gaia} Data Processing and Analysis Consortium (DPAC,\url{https://www.cosmos.esa.int/web/gaia/dpac/consortium}). Funding for the DPAC has been provided by national institutions, in particular the institutions participating in the {\it Gaia}Multilateral Agreement.
\end{acknowledgements}

\bibliographystyle{apalike}

\bibliographystyle{aa} 

\begin{thebibliography}{}
\bibliographystyle{apalike}

\bibitem[Alonso et~al., 1999]{1999A&AS..140..261A}Alonso, A., Arribas, S., and Mart{\'\i}nez-Roger, C., 1999, A\&AS, 140, 261
\bibitem[Bailer-Jones et~al., 2021]{2021AJ....161..147B}Bailer-Jones, C.~A.~L., Rybizki, J., Fouesneau, M., Demleitner, M., and Andrae, R.\ 2021, AJ, 161, 147
\bibitem[Bressan et~al., 2012 ]{2012MNRAS.427..127B}Bressan, A., Marigo, P., Girardi, L. et al.\ 2012, MNRAS, 427, 127
\bibitem[Carpenter, 2001]{2001AJ....121.2851C}Carpenter, J.~M.\ 2001, AJ, 121, 2851
\bibitem[Casagrande et~al., 2010]{2010A&A...512A..54C}Casagrande, L., Ram{\'\i}rez, I., Mel{\'e}ndez, J., Bessell, M., and
  Asplund, M.\ 2010, A\&A, 512A, 54
\bibitem[Choi et~al., 2016]{2016ApJ...823..102C}Choi, J., Dotter, A., Conroy, C. et al.\ 2016, ApJ, 823, 102
\bibitem[Coelho et~al., 2005]{2005A&A...443..735C}Coelho, P., Barbuy, B., Mel{\'e}ndez, J., Schiavon, R.~P., and  Castilho, B.~V.\ 2005, A\&A, 443, 735
\bibitem[di Benedetto, 1998]{1998A&A...339..858D}di Benedetto, G.~P.\ 1998, A\&A, 339, 858
\bibitem[Di Benedetto, 2005]{2005MNRAS.357..174D}Di Benedetto, G.~P.\ 2005, MNRAS, 357, 174
\bibitem[Dotter, 2016]{2016ApJS..222....8D}Dotter, A.\ 2016, ApJS, 222, 8
\bibitem[Eggleton and Yakut, 2017]{2017MNRAS.468.3533E}Eggleton, P.~P. and Yakut, K.\ 2017, MNRAS, 468, 3533
\bibitem[Gaia Collaboration et~al., 2022]{2022arXiv220800211G}Gaia Collaboration and Vallenari, A. and Brown, A.~G.~A. and Prusti, T. et al.\ 2022, A\&A, arXiv e-print: arXiv:2208.00211
\bibitem[Gaia Collaboration et~al., 2021 ]{2021A&A...649A...1G}Gaia Collaboration, Brown, A.~G.~A., Vallenari, A., Prusti, T. et al.\ 2021, A\&A, 649A, 1
\bibitem[Gonz{\'a}lez and Levato, 2006]{2006A&A...448..283G}Gonz{\'a}lez, J.~F. and Levato, H.\ 2006, A\&A, 448, 283
\bibitem[Gonz{\'a}lez Hern{\'a}ndez and Bonifacio, 2009]{2009A&A...497..497G}Gonz{\'a}lez Hern{\'a}ndez, J.~I. and Bonifacio, P.\ 2009, A\&A, 497, 497
\bibitem[Graczyk et al.(2022)]{2022arXiv220807257G} Graczyk, D., Pietrzy{\'n}ski, G., Galan, C., et al.\ 2022, arXiv e-print: arXiv:2208.07257
\bibitem[Graczyk et~al., 2021]{2021A&A...649A.109G}Graczyk, D., Pietrzy{\'n}ski, G., Galan, C. et al.\ 2021, A\&A, 649, 109
\bibitem[Graczyk et~al., 2018]{2018ApJ...860....1G}Graczyk, D., Pietrzy{\'n}ski, G., Thompson, I.~B. et al.\ 2018, ApJ, 860, 1G
\bibitem[Graczyk et~al., 2012]{2012ApJ...750..144G}Graczyk, D., Pietrzy{\'n}ski, G., Thompson, I.~B. et al.\ 2012, ApJ, 750, 144G
\bibitem[Gustafsson et~al., 2008]{2008A&A...486..951G}Gustafsson, B., Edvardsson, B., Eriksson, K. et al.\ 2008, A\&A, 486, 951
\bibitem[He{\l}miniak et~al., 2015]{2015MNRAS.448.1945H}He{\l}miniak, K.~G., Graczyk, D., Konacki, M. et al.\ 2015, MNRAS, 448, 1945
\bibitem[He{\l}miniak et~al., 2019]{2019MNRAS.484..451H}He{\l}miniak, K.~G., Konacki, M., Maehara, H. et al.\ 2019, MNRAS, 484, 451
\bibitem[Houdashelt et~al., 2000]{2000AJ....119.1448H}Houdashelt, M.~L., Bell, R.~A., and Sweigart, A.~V.\ 2000, AJ, 119, 1448
\bibitem[Klinglesmith and Sobieski, 1970]{1970AJ.....75..175K}Klinglesmith, D.~A. and Sobieski, S.\ 1970, AJ, 75, 175
\bibitem[Lucy, 1967]{1967ZA.....65...89L}Lucy, L.~B.\ 1967, ZA, 65, 89L
\bibitem[Masana et~al., 2006]{2006A&A...450..735M}Masana, E., Jordi, C., and Ribas, I.\ 2006, A\&A, 450, 735
\bibitem[Miglio et~al., 2012]{2012MNRAS.419.2077M}Miglio, A., Brogaard, K., Stello, D. et al.\ 2012, MNRAS, 419, 2077M
\bibitem[Paxton et~al., 2011]{MESA-I}Paxton, B., Bildsten, L., Dotter, A. et al.\ 2011, ApJS, 192, 3P
\bibitem[Paxton et~al., 2013]{MESA-II}Paxton, B., Cantiello, M., Arras, P. et al.\ 2013, ApJS, 208, 4
\bibitem[Paxton et~al., 2015]{MESA-III}Paxton, B., Marchant, P., Schwab, J.\ 2015, ApJS, 220, 15
\bibitem[Paxton et~al., 2018]{MESA-IV}Paxton, B., Schwab, J., Bauer, E.~B.\ 2018, ApJS, 234, 34
\bibitem[Pietrzy{\'n}ski et~al., 2019]{2019Natur.567..200P}Pietrzy{\'n}ski, G., Graczyk, D., Gallenne, A. et al.\ 2019, Nature, 567, 200P
\bibitem[Pietrzy{\'n}ski et~al., 2013]{2013Natur.495...76P}Pietrzy{\'n}ski, G., Graczyk, D., Gieren, W. et al.\ 2013, Nature, 495, 76P
\bibitem[Pietrzy{\'n}ski et~al., 2009]{2009ApJ...697..862P}Pietrzy{\'n}ski, G., Thompson, I.~B., Graczyk, D.\ 2009, ApJ, 697, 76P
\bibitem[Pilecki et~al., 2017]{2017ApJ...842..110P}Pilecki, B., Gieren, W., Smolec, R. et al.\ 2017, ApJ, 842, 110
\bibitem[Pilecki et~al., 2015]{2015ApJ...806...29P}Pilecki, B., Graczyk, D., Gieren, W. et al.\ 2015, ApJ, 806, 29P
\bibitem[Ram{\'\i}rez and Mel{\'e}ndez, 2005]{2005ApJ...626..465R}Ram{\'\i}rez, I. and Mel{\'e}ndez, J.\ 2005, ApJ, 626, 465
\bibitem[Reimers, 1985]{1985iue..prop.2255R}Reimers, D.\ 1985, IUE Proposal, 2255R
\bibitem[Rucinski, 1999]{1999ASPC..185...82R}Rucinski, S.\ 1999, ASPC, 185, 82R
\bibitem[Rucinski, 1992]{1992AJ....104.1968R}Rucinski, S.~M.\ 1992, AJ, 104, 1968
\bibitem[Schlafly and Finkbeiner, 2011]{2011ApJ...737..103S}Schlafly, E.~F. and Finkbeiner, D.~P.\ 2011, ApJ, 737, 103S
\bibitem[Schlegel et~al., 1998]{1998ApJ...500..525S}Schlegel, D.~J., Finkbeiner, D.~P., and Davis, M.\ 1998, ApJ, 500, 525S
\bibitem[Soszy{\'n}ski et~al., 2016]{2016AcA....66..405S}Soszy{\'n}ski, I., Pawlak, M., Pietrukowicz, P. et al.\ 2016, AcA, 66, 405S
\bibitem[Soszy{\'n}ski et~al., 2012]{2012AcA....62..219S}Soszy{\'n}ski, I., Udalski, A., Poleski, R.\ 2012, AcA, 62, 219
\bibitem[Southworth et~al., 2004]{2004MNRAS.351.1277S}Southworth, J., Maxted, P.~F.~L., and Smalley, B.\ 2004, MNRAS, 351, 1277
\bibitem[Southworth et~al., 2005]{2005MNRAS.363..529S}Southworth, J., Smalley, B., Maxted, P.~F.~L. et al.\ 2005, MNRAS, 363, 529
\bibitem[Suchomska et~al., 2019]{2019A&A...621A..93S}Suchomska, K., Graczyk, D., Pietrzy{\'n}ski, G. et al.\ 2019, A\&A, 621A, 93S
\bibitem[Suchomska et~al., 2015]{2015MNRAS.451..651S}Suchomska, K., Graczyk, D., Smolec, R. et al.\ 2015, MNRAS, 451, 651S
\bibitem[Tkachenko, 2015]{2015A&A...581A.129T}Tkachenko, A.\ 2015, A\&A, 581A, 129
\bibitem[Torres et~al., 2010]{2010A&ARv..18...67T}Torres, G., Andersen, J., and Gim{\'e}nez, A.\ 2010, A\&A Rev., 18, 67
\bibitem[Tsymbal, 1996]{1996ASPC..108..198T}Tsymbal, V.\ 1996, ASPC, 108, 198
\bibitem[Udalski, 2003]{2003AcA....53..291U}Udalski, A.\ 2003, AcA, 53, 291
\bibitem[van Hamme, 1993]{1993AJ....106.2096V}van Hamme, W.\ 1993, AJ, 106, 2096
\bibitem[Van Hamme and Wilson, 2007]{2007ApJ...661.1129V}Van Hamme, W. and Wilson, R.~E.\ 2007, ApJ, 661, 1129
\bibitem[Wilson, 1979]{1979ApJ...234.1054W}Wilson, R.~E.\ 1979, ApJ, 234, 1054
\bibitem[Wilson, 1990]{1990ApJ...356..613W}Wilson, R.~E.\ 1990, ApJ, 356, 613
\bibitem[Wilson and Devinney, 1971]{1971ApJ...166..605W}Wilson, R.~E. and Devinney, E.~J.\ 1971, ApJ, 166, 605
\end{thebibliography}
\label{lastpage}
\end{document}